\documentclass{article}

\usepackage[english]{babel}

\usepackage[letterpaper,top=2cm,bottom=2cm,left=3cm,right=3cm,marginparwidth=1.75cm]{geometry}

\usepackage{amsmath}
\usepackage{titling}
\usepackage{graphicx}
\usepackage{array}
\usepackage{booktabs}
\usepackage{geometry}
\usepackage{cellspace} 
\usepackage{longtable}
\usepackage{makecell}

\usepackage[colorlinks=true, allcolors=blue]{hyperref}
\UseRawInputEncoding
\title{\textbf{MoA is All You Need :Building LLM Research Team using Mixture of Agents} }
\author{\textbf{Sandy Chen, Leqi Zeng, Abhinav Raghunathan, Flora Huang, Terrence C. Kim}\\
Vanguard IMFS (Investment Management FinTech Strategies)}
\date{}

\begin{document}
\maketitle

\begin{abstract}
Large Language Models (LLMs) research in the financial domain is particularly complex due to the sheer number of approaches proposed in literature. Retrieval-Augmented Generation (RAG) has emerged as one of the leading methods in the sector due to its inherent groundedness and data source variability. In this work, we introduce a RAG framework called Mixture of Agents (MoA) and demonstrate its viability as a practical, customizable, and highly effective approach for scaling RAG applications. MoA is essentially a layered network of individually customized small language models \cite{hoffmann_2022_training} collaborating to answer questions and extract information. While there are many theoretical propositions for such an architecture and even a few libraries for generally applying the structure in practice, there are limited documented studies evaluating the potential of this framework considering real business constraints such as cost and speed. We find that the MoA framework, consisting of small language models \cite{hoffmann_2022_training}, produces higher quality and more grounded responses across various financial domains that are core to Vanguard’s business while simultaneously maintaining low costs.
\end{abstract}

\section{Introduction}

It is well known in the machine learning community that single-model approaches typically fall short in predictive power compared to multi-model approaches (also known as “ensemble models”). Two main reasons being:
\begin{itemize}
    \item Conclusions drawn from ensemble models are bolstered by the consensus of multiple models, each receiving slightly different inputs. This collective validation enhances the confidence in the predictive outcomes. 
    \item Ensemble models are better equipped to generalize to new information that has not been captured in the training data.
\end{itemize}

\noindent Large Language Models (LLMs) initially relied on single dense transformer approaches due to their computational complexity and the inherent risk of hallucinations. However, the research community has recently shifted its focus towards sparse ensembles of LLMs, as they offer several advantages \cite{shen_2024_learning}\cite{cheng_2023_unlocking}. These ensembles exhibited lower hallucination rates, improved output quality, and enhanced information surfacing capabilities \cite{gordon_2023_multiai}. Moreover, by arranging multiple LLMs in sequence or parallel, researchers can create intricate networks \cite{chuang_2023_simulating} that resemble the organizational structures found within real corporations. This arrangement unlocks a crucial collaborative potential, enabling LLMs to work together in a more sophisticated manner.\\

\noindent Large Language Models (LLMs) that surpass simple classification tasks and can perform actions based on information stored in databases, APIs, and other sources are known as "agents." Both individual agents and systems composed of multiple agents (often referred to as "Socratic AI," "Agentic AI," or similar terms) are extremely powerful as they can arbitrarily read and execute tasks far more efficiently than humans \cite{zeng_2022_socratic}. This capability is particularly valuable in the finance domain, where the vast amount of knowledge researchers consume is of a textual nature. For the purpose of this paper, we define a Mixture of Agents (MoA) system as an ensemble of agents, each with unique characteristics such as customized linking, prompting, and knowledge.\\

\noindent Existing literature explores ensemble LLMs primarily from a theoretical perspective, focusing on experimentation to determine if error improves or compounds in these systems. Studies have provided evidence that ensembles of LLMs can improve classification accuracy over single models \cite{li_2024_more} and collaborate through debate to solve complex problems \cite{guo_2024_large}. It is also clear that ensemble LLMs have a wide variety of potential use cases in the biomedical, financial, and even research domains \cite{chuang_2023_simulating}. The main drawback of ensemble LLMs are cost and speed - running multiple models in parallel or in sequence is a computationally costly operation that results in slow generation and high latency. \\

\begin{figure}[htbp]
\centering
\includegraphics[width=1\linewidth]{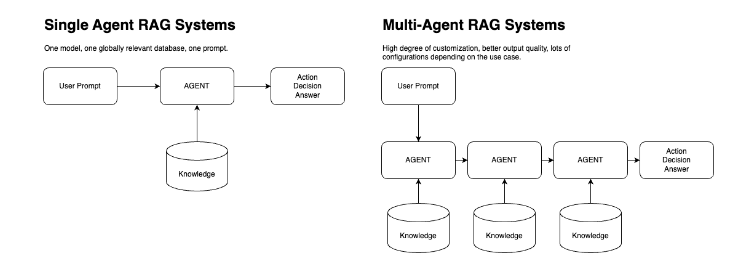}
\caption{\label{fig:moa}Single vs. Multi-Agent system configuration.}
\end{figure}

\noindent In the practical domain, research aligns more closely with intra-model approaches. Mistral AI’s groundbreaking paper on their Mixture of Experts (MoE) model \cite{jiang_2024_mixtral} seems to be motivated, at least partially, from ensemble models in traditional machine learning. Mistral’s Mixtral 8x16,a MoE model, outperformed much of the existing open-source competition due to its innovative architecture, which serves as an inspiration for our work. While MoE is model-centric, applying ensemble learning within a single model, MoA is a system-centric approach to apply ensemble learning across multiple models. OpenAI has also openly embraced the idea of ensembles. The GPT-4 model is rumored to be one of the most impactful implementations of MoE, with “GPTs” representing their active exploration of agents using GPT-4 as the foundation. Although libraries such as AIFlows, Langchain, and Microsoft Autogen enable programmatic composition of agents and LLMs, there are still very limited studies that demonstrate the viability of systems of agents when considering cost and user experience as primary factors \cite{josifoski_2024_flows}\cite{introduction}.  At Vanguard’s Investment Management Fintech Strategies (IMFS) team, we propose one of the first data points suggesting that MoA meets these constraints.

\section{Mixture of Agents (MoA)}

Mixture of Agents (MoA) is an enhanced multi-agent Retrieval-Augmented Generation (RAG) framework that supports a group of highly specialized small \cite{hoffmann_2022_training} language model agents working together in complex formations to answer questions. MoA is highly inspired by ongoing research into ensemble approaches for LLMs, including Mixture of Experts (MoE) and Socratic AI \cite{zeng_2022_socratic}\cite{yang_socraticai}. Our findings suggest that these agents operate in powerful ways that mimic organizational hierarchies, ultimately producing higher quality outputs with built-in transparency and grounding.\\

\noindent The agents that constitute the MoA system are sophisticated information gatherers, each possessing its own internal knowledge, external knowledge bases \cite{ding_2024_entgpt}, prompts, groundings, abilities, and connections with other agents. This high degree of specialization enables the overall MoA system to develop “diverse views” that converge to form a final response. More importantly, we observe that a robust MoA system consisting of small \cite{hoffmann_2022_training} language models is incredibly cost-effective. When combined with good data engineering practices, MoA can achieve speed and scale that truly rivals other methods of interacting with traditional single large language models \cite{wang_2024_mixtureofagents}. This makes MoA a suitable approach for most, if not all, enterprise use cases.

\subsection{Agent as "Junior Researcher"}

The role of the agent in the MoA framework resembles a junior researcher for investment management, but with tremendous potential. By customizing the knowledge accessible to each agent, we can develop highly diversified yet extremely “intelligent” agents that possess domain understanding and specialization.

\begin{figure}[htbp]
\centering
\includegraphics[width=1\linewidth]{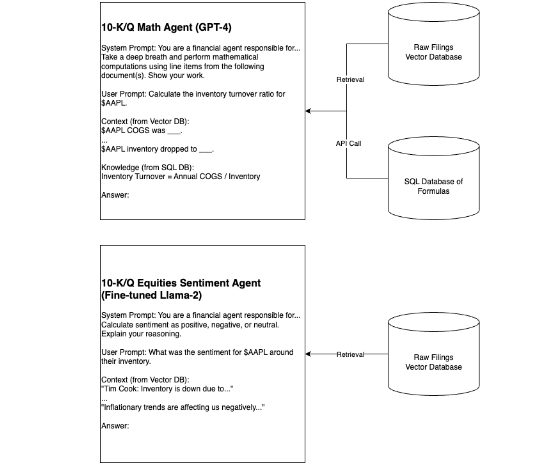}
\caption{\label{fig:fig2}Examples of hyper specialized agents with API access and knowledge.}
\end{figure}

\noindent By hyper-specializing each of these agents, they can individually achieve better results compared to a single model handling both tasks. Figure 2 illustrates an example of such agents, each with its own prompts, knowledge, instructions, fine-tuning, and model bases. In this example, the “10-K/Q Math Agent'' is a GPT-4 instance with a definitional understanding of line items and accounting terminology. It is fine-tuned and prompted specifically for mathematical tasks (“take a deep breath”). Additionally, it has RAG access to raw filings and API access to a SQL database containing analyst notes with domain-specific equations. The "10-K/Q Sentiment Agent," on the other hand, is a Llama-2 instance fine-tuned on equities sentiment classification. It has RAG access to real positive and negative statements from the company being queried and is prompted for sentiment analysis.\\

\noindent The split-agent approach offers significantly higher response quality compared to a single-model approach due to the customizability of each individual agent. These specialized agents can answer extremely nuanced and complex questions with greater accuracy and depth in an MoA system.

\subsection{Team of Junior Researcher Agents}

After agents are customized and built, it is immediately evident that for various high-level tasks, pipelines of agents can be constructed in an efficient manner to carry this task to completion. This structure is reminiscent of a “research team,” where experts with different backgrounds (i.e., agents with different customizations) collaborate to tackle a common problem. Using the same example agents from before, it is possible to pose adjacent questions to different agents to obtain more specific responses, which can then be compiled into a comprehensive answer. Figure 3 represents one possible configuration of these two agents, preceded by a “planner” that selects the questions and followed by an “aggregator” that intelligently combines the agents’ responses.

\begin{figure}[htbp]
\centering
\includegraphics[width=0.75\linewidth]{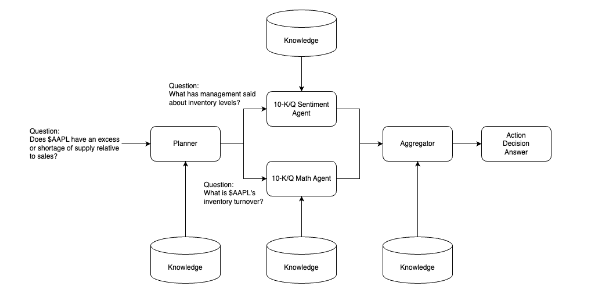}
\caption{\label{fig:fig3}Possible split of specialized agents to answer a complex question with strong response quality.}
\end{figure}

\noindent The flexibility of MoA lies in the fact that agents can be replaced by heuristics, API calls, or any other subprocess that might feed additional information into the aggregator or other agents.\\

\noindent In all these scenarios, MoA greatly benefits from its ability to maintain a high level of customizability. Since each agent effectively serves as a real-time expert on user questions, the overall action and response quality remains strong. However, it is important to note that MoA’s performance is only as good as its data and engineering capabilities. The system can potentially be allowed to grow arbitrarily complex, with reports and answers from an output potentially feeding future inputs through various different approaches currently existing and in development. At Vanguard’s IMFS team, our MoA system has scaled to analyzing tens of thousands of documents simultaneously.\\

\noindent MoA has a unique property wherein any higher-level agent responsible for summarizing or supervising the outputs of lower-level agents can discern and filter out irrelevant or inaccurate information. Interestingly, we observe that the concept of "compounding error" only occurs with a single stream of serial models and not with MoA.

\section{Results}

MoA is a property of a LLM system, unlike MoE, which is a property of LLMs. Therefore, we abstract away from the complexities of model evaluation and instead focus on higher-level results as a consequence of the system. Consistent with the findings of other researchers, we find that an interwoven network of models outperforms any single workstream. Furthermore, as the system scales and the layers of abstraction increase, both latency and potential grow. The more abstraction present, the more steps are saved for the human researchers. MoA essentially becomes increasingly efficient compared to human effort as it scales. MoA presents an extremely useful solution for those seeking to enhance existing RAG pipelines beyond the response quality of a single-model system.

\subsection{Information Surfacing \& Output Quality}

MoA enhances the information surfacing capabilities of any RAG implementation, thereby increasing the quality of the output. \\

\noindent One of the most pressing concerns regarding RAG systems is the available context window. When this value is small, the model’s coverage with respect to available data is correspondingly limited. Extensive ongoing research focuses on maximizing context windows while minimizing performance degradation \cite{liu_2023_lost}. In this regard, MoA provides an advantage over single-model systems. When using systems of agents, the effective context window of the system is significantly augmented. Instead of one model handling all available context, it can now be deliberately split among multiple expert agents. This approach allows for a higher degree of precision and reduces the probability of “lost in the middle” issues. Responses to questions with known or verifiable answers are typically well-formed and accurate. Furthermore, it is well known that model responses are extremely sensitive to their system prompts. Customizing prompts for agents based on their data source can dramatically improve output quality and insight.\\

\noindent At Vanguard, we regularly employ MoA to extract and surface insights from documents researchers refer to, often in the volume of tens of thousands of documents. We find that one of the main value propositions of MoA is the ability to drastically increase the likelihood of surfacing information that researchers may have missed or overlooked. Each agent's answers contain different contexts; for example, a 10-K filing may contain a revenue figure, while the context for this number may be found in an earnings transcript. When these diverse contexts are combined, analysts can glean circumstantial information and gain deeper levels of insight.

\subsection{Comparison with Single-Model Providers}

MoA has demonstrated better and more holistic response quality without the use of fine-tuning \cite{wang_2024_mixtureofagents}. As an experiment, major providers of single-model LLM systems (Anthropic’s Claude 3 Opus and OpenAI’s ChatGPT 4) were given two public documents: Apple’s Q1 2023 earnings transcript and 10-Q. All three models, including a rudimentary MoA system with two Mistral-7B agents (one for each document), were asked questions and graded based on the amount of vital information captured in their responses.\\

\noindent By way of example, one such question was “In \$AAPL 2023Q1 transcripts and filings, what was the context around revenue growth for the rest of the year?” The critical information that needed to be captured included:
\begin{enumerate}
    \item Revenue for Q2 2023 is projected to be similar to Q1.
    \item Gross margins are expected to be between 44-44.5\%.
    \item Macroeconomic headwinds in digital advertising and mobile gaming.
    \item Negative foreign exchange impacts.
    \item Expected stock repurchases and buybacks.
    \item Expected 65\% deferred revenue to be realized in a year.
    \item iPhone sales are expected to accelerate, while Mac/iPad sales are expected to decline.
\end{enumerate}

\begin{longtable}{@{}p{2cm} p{12cm}@{}}
    \toprule
    \textbf{Model} & \textbf{Response} \\
    \midrule
    \endfirsthead
    \toprule
    \textbf{Model} & \textbf{Response} \\
    \midrule
    \endhead
    \midrule
    \multicolumn{2}{r}{\textit{Continued on next page}} \\
    \midrule
    \endfoot
    \bottomrule
    \endlastfoot
        Claude 3 Opus (4/7)
 & Based on Apple's Q1 2023 earnings call transcript and 10-Q filing, here are 
the key points regarding their outlook for revenue growth in fiscal 2023:
\begin{enumerate}
     \item Apple is not providing specific revenue guidance for Q2 or the full year due to continued macroeconomic uncertainty. However, they did provide some directional insights.
    \item For Q2 2023, Apple expects year-over-year revenue performance to be similar to the December quarter (Q1). This represents an acceleration in underlying year-over-year business performance since Q1 benefited from an extra week compared to Q2.
    \item Foreign exchange will continue to be a headwind in Q2, with Apple 
expecting a 5 percentage point negative year-over-year impact.
    \item For iPhone, Apple expects the Q2 year-over-year revenue performance to 
accelerate relative to the Q1 year-over-year revenue decline.
    \item For Mac and iPad, Apple expects Q2 revenue for both categories to decline double-digits year-over-year due to challenging compares and macroeconomic headwinds.
    \item Services revenue is expected to grow year-over-year in Q2 while 
continuing to face macroeconomic headwinds in areas like digital advertising and mobile gaming.
 \end{enumerate}    
 Overall, while Apple faced short-term revenue declines due to external factors, the company's strategic focus on services and emerging markets, along with its strong product ecosystem, were seen as key drivers for potential growth throughout the rest of the year.\\
 GPT-4o (2/7) & 
 \begin{itemize}
     \item In Apple's Q1 2023 earnings report, the company faced several challenges that impacted its revenue growth outlook for the rest of the year. Notably, Apple experienced a 5\% year-over-year decline in revenue, amounting to \$117.2 billion, primarily due to supply chain shortages and weaker iPhone sales. This decline was significant given that Q1 typically includes the holiday shopping season, which is crucial for Apple's financial performance.
     \item  Tim Cook, Apple's CEO, highlighted that despite these challenges, the company maintained a strong product lineup and focused on long-term growth. He mentioned that Apple's installed base of active devices surpassed 2 billion, reflecting high customer loyalty and satisfaction across all product categories.
    \item Looking forward, Apple CFO Luca Maestri emphasized the company's resilience in the face of a difficult macroeconomic environment. He noted that Apple's services sector set an all-time revenue record of \$20.9 billion, showing robust growth even amidst broader economic pressures. Additionally, Apple saw strong performance in emerging markets, with significant growth in regions like South Asia, India, Latin America, and the Middle East.
 \end{itemize}\\
        ChatGPT 4(3/7)
 & In the 2023 Q1 filings and transcripts for Apple, the discussion around 
revenue growth for the rest of the year highlighted several challenges and 
strategic focuses. Here are the key points:
\begin{enumerate}
    \item  **Revenue Challenges and Achievements**: Apple reported a revenue of 
\$117.2 billion for the December quarter, marking a decrease of 5\% year over
year. Despite this decline, they achieved all-time revenue records in multiple 
markets globally. The factors affecting revenue included foreign exchange 
headwinds, COVID-19 related supply chain disruptions, particularly with 
iPhone 14 Pro models, and a challenging macroeconomic environment. 
However, production had returned to desired levels by the end of the quarter.
    \item **Strategic Initiatives and Product Performance**: Apple's CEO Tim Cook 
emphasized the company's continuous investment in innovation and their 
strategic initiatives across various product categories. iPhone revenue was 
slightly down but flat on a constant currency basis. The Mac and iPad lines 
showed robust performances, with the Mac maintaining strong customer 
upgrade activity and the iPad growing due to a better supply situation 
compared to the previous year.
    \item **Services Growth**: Apple's services segment reached an all-time 
revenue record of \$20.8 billion, up 6\% year over year. This growth was driven by strong performance in cloud services, payment services, and an increase in paid subscriptions. Apple now hosts over 935 million paid subscriptions across its services.
    \item **Forward Outlook**: While specific revenue guidance was not provided 
due to ongoing macroeconomic uncertainties and potential COVID-19 
impacts, the company shared directional insights suggesting that the year-
over-year revenue performance in the March quarter would be similar to the December quarter. They expected continued foreign exchange headwinds but also saw potential revenue growth in services.
\end{enumerate}\\

MoA Agent1 (5/7)
& Apple's revenue for Q2 FY23 is expected to be similar to that of Q1 FY23, with a negative year-over-year impact of nearly 4 percentage points due to foreign exchange. Services revenue growth is also expected to be similar to Q1 FY23, while facing macroeconomic headwinds in areas such as digital advertising and mobile gaming. Gross margin is expected to be between 44\% and 44.5\%. The company expects to continue to manage for the long term and invest in innovation and product development, while closely managing spend. Despite the challenges, Apple continues to see strong growth in its installed base of over 2 billion active devices and growing customer engagement with its services. The company also plans to return \$90 billion to shareholders through share repurchases and dividends, maintaining its goal of getting to net cash neutral over time.\\

MoA Agent2 (6/7)
& The context around revenue growth for the rest of the year in Apple Inc.'s (\$AAPL) 2023Q1 transcripts and filings indicates that the Rest of Asia Pacific net sales increase during the second quarter and first six months of 2023 was primarily driven by higher net sales of iPhone, partially offset by lower net sales of Mac. However, the Company also mentions that gross margins are subject to volatility and downward pressure due to various factors, as discussed in the 2022 Form 10-K under the heading "Risk Factors." The Company expects 65\% of total deferred revenue to be realized in less than a year, 26\% within one-to-two years, 7\% within two-to-three years, and 2\% in greater than three years. There have been no material changes to the Company’s market risk during the first six months of 2023. The Company's fiscal years 2023 and 2022 span 53 and 52 weeks, respectively. New product and service introductions can significantly impact net sales, cost of sales, and operating expenses.\\

MoA Aggregator (7/7)
& According to Apple's Q1 FY23 transcripts and filings, the company expects its revenue for Q2 FY23 to be similar to that of Q1 FY23, with a negative year-over-year impact of nearly 4 percentage points due to foreign exchange. Services revenue growth is also expected to be similar to Q1 FY23, while facing macroeconomic headwinds in areas such as digital advertising and mobile gaming. Despite these challenges, Apple continues to see strong growth in its installed base of over 2 billion active devices and growing customer engagement with its services. The company plans to return \$90 billion to shareholders through share repurchases and dividends, maintaining its goal of getting to net cash neutral over time. Additionally, the company mentions that gross margins are subject to volatility and downward pressure due to various factors, and 65\% of total deferred revenue is expected to be realized in less than a year.\\
        
    \bottomrule

    \label{tab:model_response}
\end{longtable}

\noindent The analysis clearly demonstrates that a MoA system consisting of two Mistral-based agents (each with 7B parameters) competes effectively with much larger and more expensive systems such as ChatGPT 4, GPT-4o and Claude 3 Opus. Interestingly, MoA and Claude excel at reporting on expected or future states, as the question requires, in contrast to ChatGPT 4. Responses from each agent can also be displayed to the user for transparency. Most of the responses to other questions asked followed closely with the example laid out above.

\subsection{Cost \& Scale}

The beauty of the current MoA system lies in its cost effectiveness and simplicity. In its simplest form, MoA can be performed with the same model and endpoint, activated as many times as necessary to perform inference through the various layers. For enterprises with cloud-based compute resources and endpoints priced based on uptime, such as Amazon Sagemaker or Microsoft Azure, there is no significant difference in overall cost between MoA and single-model systems of similar size.\\

\noindent The drawback of MoA is its higher demand on concurrent inference. When scaling, single-model systems can support more users because each user accesses only one endpoint. In contrast, MoA requires at least two endpoints per user, and this number can grow arbitrarily large. However, this flexibility allows for the customization of the agent configuration within the system based on budget and use case. Vanguard IMFS’s own MoA system has a significantly lower cost compared to most third-party RAG providers, such as Arcus and Databricks, with a total run cost of under \$8,000 per month processing a team of researchers’ queries. As for speed, Vanguard IMFS’s MoA system, which includes pre- and post-operations such as tokenization, retrieval, and hallucination catching, is capable of searching and surfacing information from over 30,000 documents in under 60 seconds using two layers of agents. The latency penalty for implementing MoA is approximately 4.07x, or 2.24x when running inference in parallel.In comparison, our original single-model system was capable of performing the same operation in under three seconds. These results, summarized in Table 2, were obtained using a rudimentary MoA with two layers: three context-accepting agents in layer one and one aggregator in layer two.

\begin{table}[htbp]
\centering

\begin{tabular}{|p{4cm}|p{3cm}|p{2cm}|p{4cm}|}
 \hline
 \textbf{Metric} & \textbf{Single-Model Systems} & \textbf{MoA} & \makecell{\textbf{MoA} \\ \textbf{(Parallel Inference)}} \\ [0.5ex] 
 \hline\hline
 Max Concurrent Users & Around 20  & 11 & 11 \\ 
 \hline
 Total Compute Cost per Month & \$5,000-\$8,000 & \$5,000-\$8,000 & \$5,000-\$8,000 \\
 \hline
 Average Response Speed & 2.9974s & 12.3334s & 6.8626s \\
 \hline
 Average Latency Penalty & - & 4.07x & 2.24x \\
 \hline
 Average Passages Considered & 30 & 90 & 90\\
 \hline
 Average Context Window Improvement & - & 3.00x & 3.00x \\ [0.5ex] 
 \hline 

\end{tabular}
\caption{Summary of speed and context window differences between single-model, MoA, and optimized MoA architectures.}
\end{table}

\noindent Based on this and other similar analyses, we conclude that the speed and context window improvement of MoA scales linearly with the number of models used in the system. In the above table, we implemented a four-model MoA, consisting of three agents accepting contexts, one aggregator. The total inference time increases by 4x without parallelization, and the context window increases by 3x as a result. \\

\noindent MoA is an efficient system that maximizes the benefits of RAG while still meeting cost and scalability constraints in practice. If an enterprise can create and deploy a single-model system, it can also deploy MoA.

\subsection{Permanence}

As a framework, MoA is a robust system that maintains an edge over traditional single-model LLM systems. At Vanguard, we have supported the hypothesis that smaller language models \cite{hoffmann_2022_training} are the present and future when it comes to highly efficient and accurate outcomes. MoA is an extension of this hypothesis, as it has allowed us to operate at a fraction of the cost by utilizing open-weight, sub-10B parameter models. With most of the language modeling community arriving at similar conclusions, we believe in MoA’s permanence to become an industry standard.

\subsection{Transparency}

Since the responses from each agent serve as an input to the final aggregator, each output can be regularly displayed to the user and evaluated for missteps or hallucinations. At its core, MoA is a variant of an advanced RAG system and, therefore, retains all of its transparency and grounding properties.\\

\noindent However, there are cases where the final output from the MoA system is not as relevant or impactful as an output from one of the constituent agents. In such situations, it is a straightforward task to present the output from each agent to the user along with the final output, allowing them to make their own judgment.

\begin{figure}[htbp]
\centering
\includegraphics[width=0.75\linewidth]{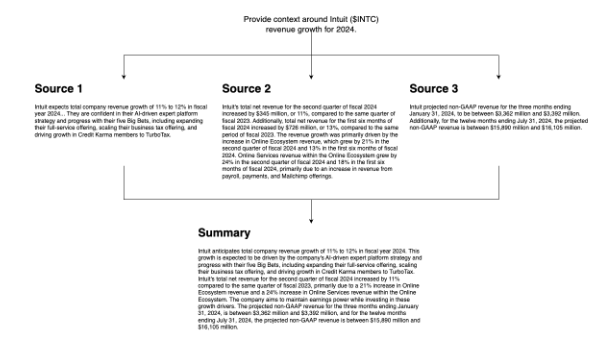}
\caption{\label{fig:fig4}Example output of MoA with Mistral v0.2 as the agent model. Each agent has its own output that can be used to verify the summary.}
\end{figure}

\noindent At Vanguard, we have invested a substantial amount of time in developing safeguards to limit the hallucination tendency of the models within the MoA system. One of the hardest tasks was to teach the models to say “I don’t know” when the model did not have the relevant dataset to answer a specific question.These safeguards range from heuristics-based checks to more complex embedding comparisons, ensuring the reliability and accuracy of the generated outputs.

\section{Conclusion \& Future Plans}

By comparing cost, output quality, transparency, and various other characteristics of LLM systems, we conclude that MoA using small language models should be the de facto standard for enterprise-grade RAG pipelines.\\

\noindent It is important to note that this analysis was conducted using a specific technology stack consisting of Amazon AWS. Performance may be significantly improved by employing more efficient cost-per-token providers such as Fireworks AI or Groq, which may also offer faster inference times and better scalability. With improved performance, the delta between MoA and single LLM systems decreases substantially. As MoA's output quality surpasses that of single LLM systems, it potentially becomes \textit{strictly better}.

\bibliographystyle{ieeetr}
\bibliography{ref}

\end{document}